\documentclass[aps,pra,longbibliography, nofootinbib, reprint, twocolumn,  amsmath, amssymb, showpacs,  floatfix, superscriptaddress]{revtex4-1}% draft twocolumn preprint,
\usepackage{amsmath,graphics, graphicx, bm, natbib, color}
\newcommand{\rem}[1]{}

\newcommand{\ud}{\mathrm{d}}

\begin{document}

\author{Wijnand Broer}
\altaffiliation[Present address: ]{Institut f\"ur Physik, Universit\"at Rostock, Albert-Einstein-Stra{\ss}e 23, 18059 Rostock, Germany}
\affiliation{Zernike Institute for Advanced Materials, University of Groningen, \\
Nijenborgh 4, 9747 AG Groningen, the Netherlands}

\author{Holger Waalkens}
\affiliation{Johann Bernoulli Institute for Mathematics and Computer Science, University of Groningen, \\
Nijenborgh 9, 9747 AG Groningen, the Netherlands}

\author{Vitaly B. Svetovoy} 
\affiliation{MESA\textsuperscript{+} Institute for Nanotechnology, University of Twente, \\
P.O. Box 217, 7500 AE Enschede, the Netherlands}

\author{Jasper Knoester}
\affiliation{Zernike Institute for Advanced Materials, University of Groningen, \\
Nijenborgh 4, 9747 AG Groningen, the Netherlands}

\author{George Palasantzas}
\affiliation{Zernike Institute for Advanced Materials, University of Groningen, \\
Nijenborgh 4, 9747 AG Groningen, the Netherlands}

\title{Nonlinear actuation dynamics of driven Casimir oscillators with rough surfaces}

\date{\today}

\begin{abstract}
At separations below 100 nm, Casimir-Lifshitz forces strongly influence the actuation dynamics of microelectromechanical systems (MEMS) in dry vacuum conditions. For a micron size plate oscillating near a surface, which  mimics a frequently used setup in experiments with
MEMS, we show that the roughness of the surfaces significantly influences the qualitative dynamics of the oscillator. Via a combination of analytical and numerical methods, it is shown that  surface roughness leads to a clear increase of initial conditions associated with chaotic motion, that eventually lead to stiction between the surfaces. Since stiction leads to malfunction of MEMS oscillators, our results are of central interest for the design of microdevices. Moreover, it is of significance for fundamentally motivated experiments performed with MEMS.

\end{abstract}
% PACS, the Physics and Astronomy Classification Scheme. 

\maketitle
\section{Introduction}
Casimir(-Lifshitz) forces are electromagnetic dispersion interactions  between neutral surfaces without permanent dipoles. These forces arise from quantum mechanical  and thermal fluctuations \cite{Casimir48,*Lifshitz55,*Lifshitz61,  Johnson2011,  Miri2008}. These dispersion forces are expected to become significant as components of MEMS enter sub-micrometer separations \cite{ CapassoReview2007, *Rodriguez2015, BuksEPL2001, Serry95, *Serry98, DelRio2005,  Esquivel2009}. The small scales at which MEM engineering is now conducted have revived interest in the Casimir force since devices such as vibration sensors and switches are made with parts that are just a few micrometers in size. They have the right size for the Casimir force to play a role: the surface areas are sufficiently big and the separations are sufficiently small for the force to draw components together and lock them tight, an effect called \emph{stiction}.  Whereas electrostatic forces can be eliminated by reducing the voltage between the surfaces, and the influence of hydrodynamic and capillary forces can be avoided by letting the device operate in a clean, dry environment, the Casimir force cannot be excluded. Unlike the other surface forces, the Casimir force can hence impose a principal limitations on MEMS applications.

At separation distances larger than 100 nm, the spatial gradient of the Casimir force can be measured very precisely with a MEMS oscillator within a linearization approximation \cite{Decca2005}.  However, at separations below 100 nm, the nonlinearity of the Casimir force has been experimentally demonstrated to have a qualitative effect on the motion of MEM systems \cite{Chan2001}. At these short separation ranges,  the Casimir force is in particular large enough to be a formidable obstacle to achieving stable actuation. In such a case, both the influence of the permittivities of the materials \cite{Lifshitz61,  Johnson2011} and that of surface roughness \cite{PeterRoughness, VitalyReview2015} must be taken into account in order to come to a realistic evaluation of the Casimir force. This is crucial for further understanding and controlling the actuation dynamics of the system in order to prevent stiction.

In this paper we study the nonlinear actuation dynamics of a damped driven Casimir oscillator including surface roughness and optical response of interacting bodies in a realistic way. This allows for the modeling of realistic MEMS oscillators. We show that surface roughness leads to an increase of chaotic dynamics that would result in stiction. 

The paper is organized as follows: after the introduction, the model of the MEM system is described. Next, the case of a conservative system is briefly revisited to proceed to its generalization, a driven oscillator. Finally, the results are summarized in the concluding section.

% --------------------------------------------------------------------------------------------

%{\em Model.---}
\section{Model}
\begin{figure}[htbp]
	\centering
		\includegraphics[width=0.48\textwidth]{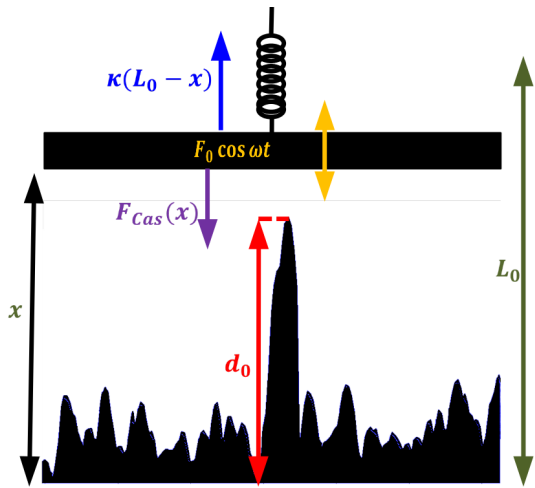}
	\caption{  Schematic of the system to clarify the meaning of the parameters. Energy dissipation and gain are allowed through damping and driving, respectively.}
	\label{fig:fig51}
\end{figure}

We model a MEMS oscillator as a classical mass-spring system, an approach well established in the study of MEMS \cite{PeleskoBernstein}. The spring models a cantilever attached to a mass of which the surface is separated less than 100 nm from another surface (see Fig.~\ref{fig:fig51}). The motion of this mass is actuated by external periodic forcing. The governing equation is of the form
\begin{equation}\label{eq:EOM}
m\ddot{x}= \kappa(L_0-x)-F_{\text{Cas}}(x) -\epsilon\gamma \dot{x} +\epsilon F_0\cos\omega t.
\end{equation}
Here $F_{\text{Cas}}(x)$ denotes the Casimir force between  two (possibly rough) parallel plates \cite{BroerEPL2011, *BroerPRB2012}, $\kappa$ is the spring constant, $L_0$  is the equilibrium distance in the case where the forces can be neglected (i.e. $F_0=0$ and $F_{Cas}(x)=0$) which is the characteristic length scale of the problem, $m$ denotes the effective mass which is determined by  the natural frequency $\omega_0\equiv\sqrt{\kappa/m}$, $\gamma$ is the friction constant, and $F_0$ is the amplitude of the periodic driving force, whose frequency is denoted by $\omega$. Typical values of the friction and driving forces are relatively small \cite{Garcia2002}. This is indicated formally by the coefficient $\epsilon=1$, which has merely an indicative character.

Throughout this paper, we choose the actuation parameter values to be $\kappa=0.5$ N/m,  $L_0=100$ nm,
and $ \omega_0 = 2\pi\cdot300$ krad/s. With the proper initial conditions, the spring constant is large enough to prevent stiction for the conservative system ($\epsilon=0$ in Eq.~\eqref{eq:EOM}) even in the rough case \cite{Broer2013}. The other actuation parameters in Eq.~\eqref{eq:EOM}, $\gamma$, $F_0$, and $\omega$, will be varied in the bifurcation analysis presented here.
 The rough surfaces are characterized by the distance upon contact $d_0$, (Fig.~\ref{fig:fig51}) which is typically about 4 to 5 times the root-mean-square roughness \cite{d0paper}. It is defined as the height of the highest asperity within a realization of an interaction area. Since $d_0$ redefines the minimum separation, the coordinate of the oscillator satisfies $x>d_0\geq0$. The case $d_0=0$ corresponds to flat surfaces. {The rough surface considered here has an r.m.s. roughness of 10.1 nm and a contact distance of $d_0= 50.8$ nm \cite{d0paper}.} 

The Casimir force is computed for two parallel plates each of size  $10\times10\mu$m$^2$. Ellipsometry data of gold films are used as input for the force calculations, which is required for a quantitative evaluation of the Casimir force \cite{PeterOptical}. In order to account for the effect of surface roughness, the results of the model from Ref.~\onlinecite{BroerEPL2011, *BroerPRB2012} are used.

Earlier works  on the actuation dynamics of MEMS under the influence of Casimir forces usually concern  conservative systems  \cite{Serry95, *Serry98, Zhao99,Esquivel2009,  Broer2013} ($\epsilon=0$ in Eq.~\eqref{eq:EOM}) or autonomous systems with damping \cite{Mehdi2013, *Mehdi2015} ($F_0=0$ in 
Eq.~\eqref{eq:EOM}). The general non-autonomous case, which is closer to an experimental MEMS oscillator setup, has been tackled analytically for ideal metals \cite{Emig2007}, as well as using an expansion of the Casimir force in the oscillator's coordinates \cite{Chan2001,Bonnin2013}. The higher order terms in such polynomial expansions give rise to additional zeros of the conservative force equation, which do not correspond to physical equilibria \cite{Bonnin2013}.  
Our approach does not rely on any such approximations, and includes the Casimir force at submicron-scale separations in an experimentally relevant way \cite{PeterRoughness, BroerEPL2011, *BroerPRB2012}.  We are not aware of other theoretical work that takes the optical response  and surface roughness into account for forced Casimir oscillators.

The friction coefficient $\gamma$ may also be written as $\gamma=m\omega_0/Q$, where $Q$ denotes the quality factor, which typically takes values of the order $10^2 - 10^3$ \cite{Mehdi2013, *Mehdi2015}. It is assumed here that the MEM system operates in clean and dry conditions: only intrinsic energy dissipation \cite{Garcia2002, Mehdi2013, *Mehdi2015}, where some of the kinetic energy of the oscillator is converted into heat is considered here. Capillary and hydrodynamic forces can be ignored. {The value of the driving amplitude $F_0$ considered here is typically of the order of several nN. In this range of values thermal noise is also negligible at room temperature \cite{Rugar2001}.}

% --------------------------------------------------------------------------------------------

%{\em The conservative oscillator  ($\epsilon=0$):---}
\section{Conservative versus Driven oscillator}
\subsection{The conservative oscillator  ($\epsilon=0$)}

\begin{figure*}[hptb]
	\centering
		\centerline{\includegraphics[width=0.48\textwidth]{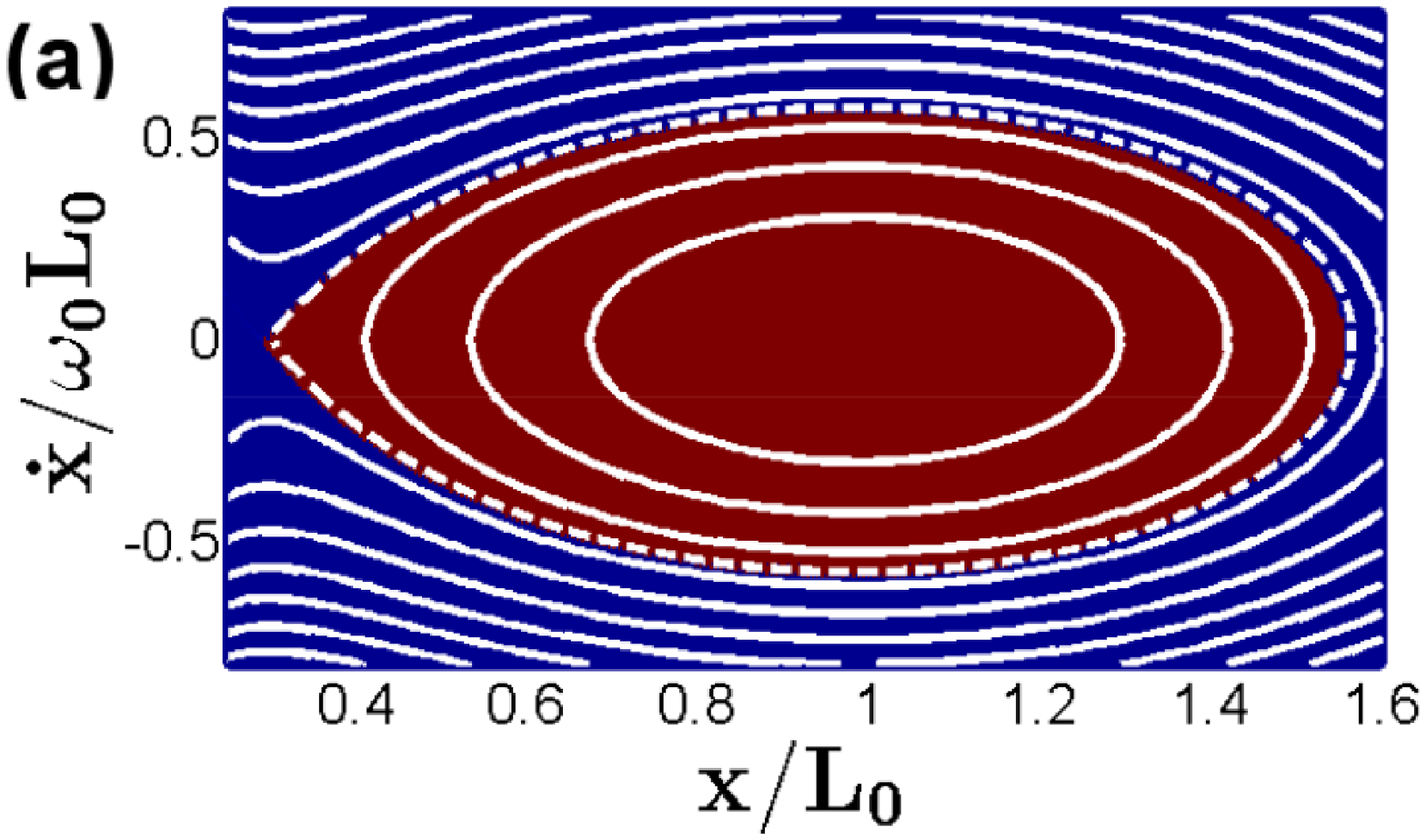}\includegraphics[width=0.48\textwidth]{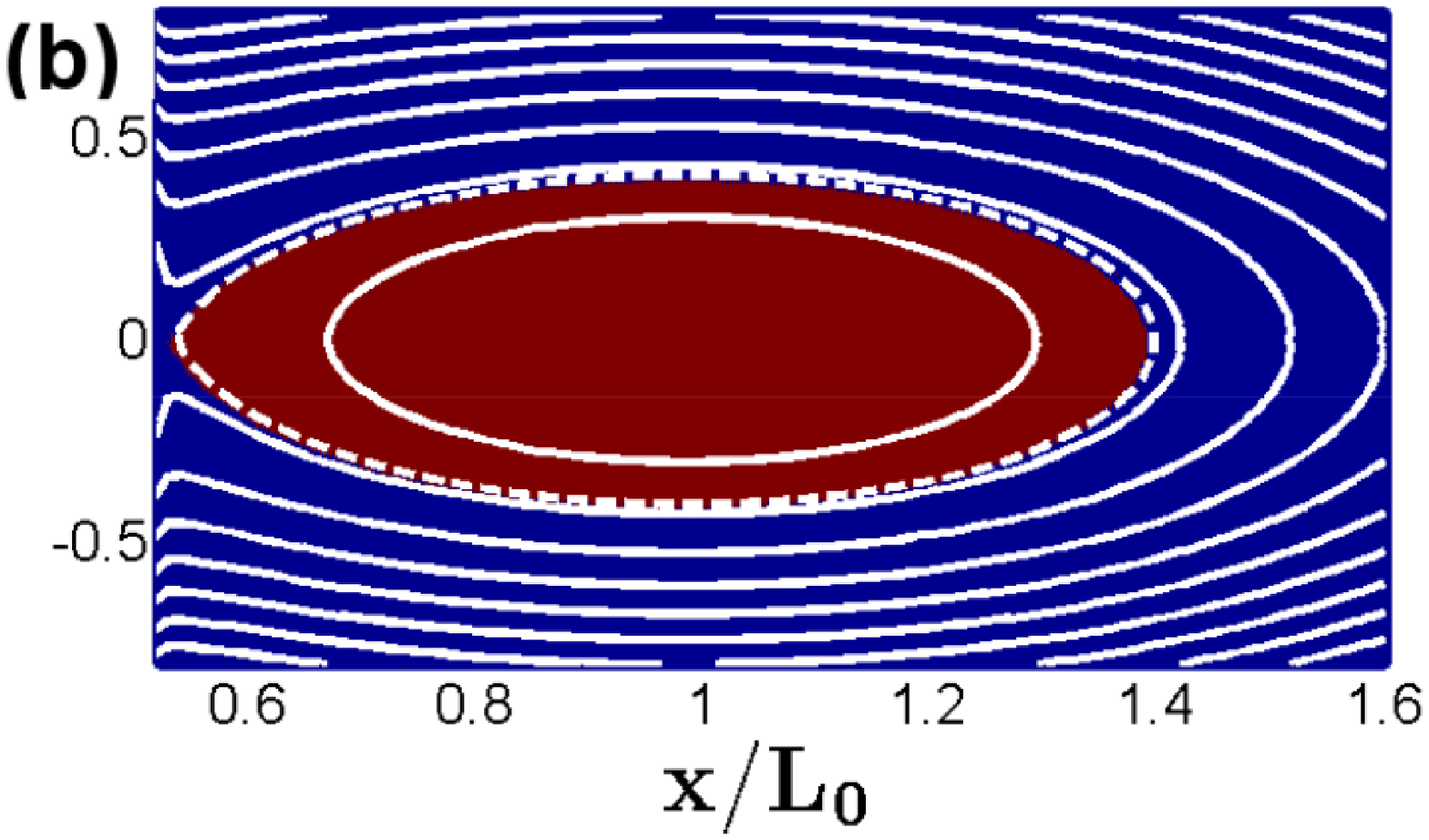}}
		
		\caption{ Grid of 300 $\times$ 300 initial conditions for the system in Eq. \eqref{eq:EOM} for $\epsilon=0$. The red region shows initial conditions for which stable actuation occurs for at least 100  natural periods $2\pi/\omega_0$. In the blue region, stiction occurs within one natural period. Panel (a) shows the result for a flat surface, and panel (b) represents a rough surface. The red region shows for which initial conditions stiction is avoided. For initial conditions in the blue region, stiction occurs within one natural period $2\pi/\omega_0$. The dashed (white) line in each figure shows the homoclinic orbit of the conservative system. The continuous (white) lines are energy level curves. }
		\label{fig:eps0}
\end{figure*}

In the conservative case, a stable center equilibrium is accompanied by an unstable saddle if the spring constant $\kappa$ is large enough \cite{Broer2013}.  The orbit in phase space which connects the saddle asymptotically to itself as time goes to $\pm \infty$ is known as the \emph{homoclinic orbit}. This orbit acts as a \emph{separatrix}, i.e. it separates qualitatively different solutions: 
in the region enclosed by the homoclinic solution we find  periodic oscillations  about the stable center corresponding to stable actuation whereas in the outer region  there is no periodic motion and every solution will lead to stiction. This is illustrated by the (white) lines in Figs. \ref{fig:eps0} (a) and (b): the homoclinic orbit is indicated by the dashed (white) line. Inside it, all curves are closed, which corresponds to periodic solutions. Outside the dashed (white) line, the curves do not return to their initial position, which indicates the absence of periodic solutions for such initial conditions. Such curves, which form the so called  \emph{phase portrait}, can be obtained in at least two ways for a conservative oscillator: firstly, by directly numerically solving Eq. \eqref{eq:EOM} for $\epsilon=0$ with e.g. the Runge-Kutta algorithm. Secondly, one can plot different level curves for different values of the (constant) energy. Both methods produce identical results indeed: Eq. \eqref{eq:EOM} for $\epsilon=0$ has been integrated on a grid of 300$\times$300 initial conditions until stiction occurs. The red region  in Fig.~\ref{fig:eps0} corresponds to initial conditions leading to stable actuation (for at least 100 natural periods $2\pi/\omega_0$), whereas an initial condition in the blue region leads to stiction within one natural period.

Physically, the presence of periodic solutions corresponds to stable actuation, whereas their absence indicates stiction. Other types of solutions do not exist in the conservative case. The fact that the homoclinic orbit strictly demarcates these qualitatively different solutions precludes the possibility of \emph{chaotic motion}, i.e. the physically observable phenomenon that the motion sensitively depends on its initial conditions \cite{GuckenheimerBook}. In other words, a chaotically moving oscillator can have qualitatively different solutions for an arbitrarily small difference in initial conditions. The case of a conservative oscillator provides an important reference for our study of a driven oscillator, since the latter will be treated as a perturbative correction of the former.

\subsection{ The driven oscillator ($\epsilon=1$)}
The explicit time dependence of the driven oscillator adds one dimension to the state space compared to the conservative case. This opens the way for the occurrence of chaotic motion. As the driving force  is small compared to the other forces, the source of the chaotic motion in the present case is the splitting of the separatrix of the conservative system, which by the  Smale-Birkhoff homoclinic theorem \cite{GuckenheimerBook} implies the occurrence of chaotic motion. 
In the first-order approximation in $\epsilon$ of the non-conservative system, the question of whether the separatrix splits can be answered in terms of the so called \emph{Melnikov function}  \cite{GuckenheimerBook} which in this case is given by
\begin{equation}\label{eq:Melnikov1}
M(t_0)=\int\limits_{-\infty}^{\infty}  \dot{x}_h(t)  [-\frac{\gamma\omega_0L_0}{F_0} \dot{x}_h(t) +  \cos \omega (t+t_0)] \, \ud t,
\end{equation}
where $x_h$ denotes the homoclinic solution.  Note that $M(t_0)$ is periodic in $t_0$ with period $2\pi/\omega$. In fact the separatrix splits if $M(t_0)$ has simple zeros, i.e. $M(t_0)=0$ and $ M'(t_0)\neq0$ for some (and due to the periodicity infinitely many) $t_0$. If $M(t_0)$ has no zeros,  the separatrices will not intersect and the motion will not be chaotic. The condition of non-simple zeros, i.e. $M(t_0)=0$ and $ M'(t_0)=0$, gives the threshold condition for chaotic motion \cite{GuckenheimerBook}. {Note that the Melnikov function is dimensionless in Eq. \eqref{eq:Melnikov1}: $t$ is expressed in units of $2\pi/\omega_0$, and $x_h$ has units of $L_0$. However, the choice of units has no bearing on the existence or the nature of the zeros of $M(t_0)$.}

In the present case, where the equation of motion has the form \eqref{eq:EOM}, the Melnikov function can be computed as \cite{Ling87}
\begin{equation}\label{eq:final_Melnikov}
M(t_0)= -\alpha \langle\dot{x}_{\text{h}}^2\rangle+A(\omega)\cos(\omega t_0+\varphi(\omega)),
\end{equation}
where the triangular brackets denote the time average with respect to a uniform distribution, i.e.  
\begin{displaymath}
\langle{}f\rangle\equiv\int\limits_{-\infty}^{\infty}f(t)\ud t,
\end{displaymath}
for an integrable function $f(t)$, and $\alpha\equiv\gamma\omega_0 L_0/F_0$. The term $A(\omega)\cos(\omega{}t_0+\varphi(\omega))$ corresponds to the real part of the Fourier transform of  $\dot{x}_{\text{h}}(t)$. (The functions $A(\omega)$ and $\varphi(\omega)$ are obtained from a polar representation of the latter.) It should be stressed that the expressions of Eqs. \eqref{eq:Melnikov1} and \eqref{eq:final_Melnikov} are applicable only to the case of additive perturbative force as in Eq.~\eqref{eq:EOM}, where $F_{\text{Cas}}(x)$ may be replaced by another nonlinear function of $x$ \cite{Ling87}.

From Eq.~\eqref{eq:final_Melnikov} we see that the threshold condition for chaotic motion, i.e. the presence of nonsimple zeros, only depends on the ratio of  $\gamma$ and $F_0$ and
it is unaffected by the phase $\varphi(\omega)$, i.e. only the amplitude $A(\omega)$  determines for which values of $\alpha$ and $\omega$ the threshold condition is met. This allows us to write the threshold condition as
\begin{equation}\label{eq:threshold}
\alpha = \frac{A(\omega)}{\langle\dot{x}_{\text{h}}^2\rangle},
\end{equation}
where the amplitude  $A(\omega)$ can be obtained by taking the absolute value of the Hilbert transform performed on $A(\omega)\cos(\omega t_0+\varphi(\omega))$. The Hilbert transform of a function $u(\tau)$ is defined as
\begin{equation}\label{eq:Hilbert}
 \mathcal{H}[u(\tau)](t) \equiv \frac{1}{\pi}P\int\limits_{-\infty}^{\infty}\ud\tau\frac{u(\tau)}{t-\tau},
\end{equation}
where $P$ denotes the principal value. Eq. \eqref{eq:Hilbert} is a convolution in the same domain as the function $u(\tau)$. The Hilbert transform is a commonly used technique in signal analysis to obtain the envelope of  an analytic signal \cite{BracewellBook}. Its connection to the Kramers-Kronig relations makes it also of particular interest to the field of optics of continuous media \cite{NussenzveigBook}.

Since $A(\omega)\cos(\omega{}t_0+\phi(\omega))$ is the real part of the Fourier transform of $\dot{x}_\text{h}(t)$, the threshold condition in Eq.~\eqref{eq:threshold} can be written explicitly in terms of the homoclinic solution of Eq. \eqref{eq:EOM}  with $\epsilon=0$ as:
\begin{equation}\label{eq:threshold_final}
 \alpha = \frac{1}{\langle\dot{x}^2_{\text{h}}\rangle}{|}\mathcal{H}[\text{Re}(\mathcal{F}[\dot{x}_\text{h}(t)])](\omega){|},
\end{equation}
where $\mathcal{F}$ denotes the Fourier transform. From Eq. \eqref{eq:threshold_final} it can be concluded that once the homoclinic solution of the conservative equation of motion is known, a statement can be made about the zeros of the Melnikov function. Hence $\dot{x}_\text{h}(t)$ also contains the information about the occurrence of chaotic motion for the driven oscillator. This also implies that the separation range of interest for the qualitative behavior of the solutions is completely determined by the homoclinic orbit of the conservative system.

\begin{figure}[htbp]
	\centering
		\includegraphics[width=0.48\textwidth]{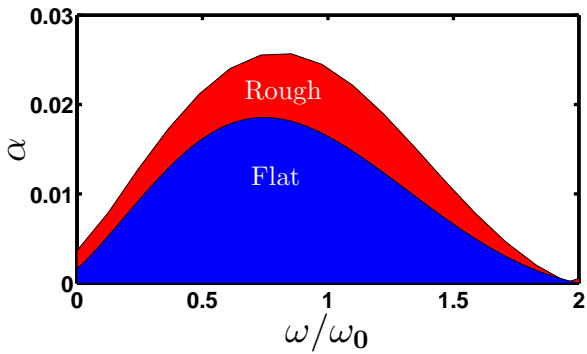}
	\caption{ The threshold curves \eqref{eq:threshold_final} for a rough and a flat surface.}
	\label{fig:threshold}
\end{figure}
Figure~\ref{fig:threshold} shows the threshold curve \eqref{eq:threshold_final} in the $(\omega,\alpha)$-plane for a flat surface and for a rough surface \cite{d0paper}. For large values of $\alpha$, the friction dominates the driving, leading to regular motion that asymptotically approaches the stable periodic orbit associated with the stable equilibrium of the conservative system. For parameter values below the curve in Fig.~\ref{fig:threshold}, the splitting of the separatrix and hence chaotic motion occurs.

In order to show the dynamical implication of the separatrix splitting we compute for each initial condition $(x(0),\dot{x}(0))$ the time the corresponding solution of Eq.~\eqref{eq:EOM} leads to stiction and plot the contours of the ``survival time'' in the $(x,\dot{x})$-plane. More specifically, we numerically integrate solutions for a uniform grid of 300$\times$300 initial conditions for six different combinations of roughness and actuation parameter values until stiction occurs or a maximum time of 100 periods of the forcing is reached. We note that the validity of our numerical procedure has been confirmed  for the Duffing Oscillator (see supplemental material \cite{BroerSupplemental2015} at [URL will be inserted by publisher] for similar plots for a Duffing oscillator). 

\begin{figure*}[hptb]
	
		\centerline{\includegraphics[width=0.48\textwidth]{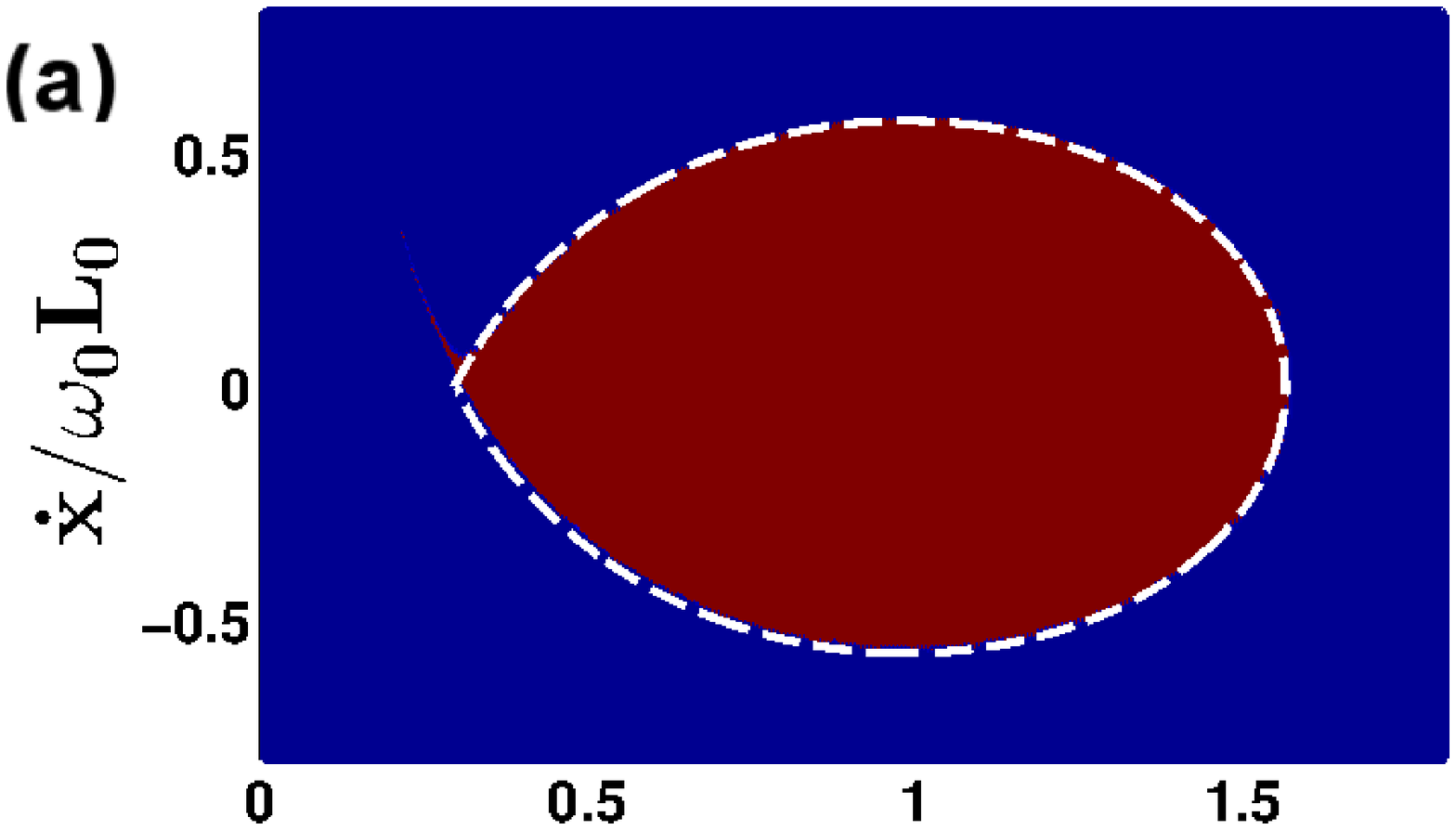}
		\includegraphics[width=0.48\textwidth]{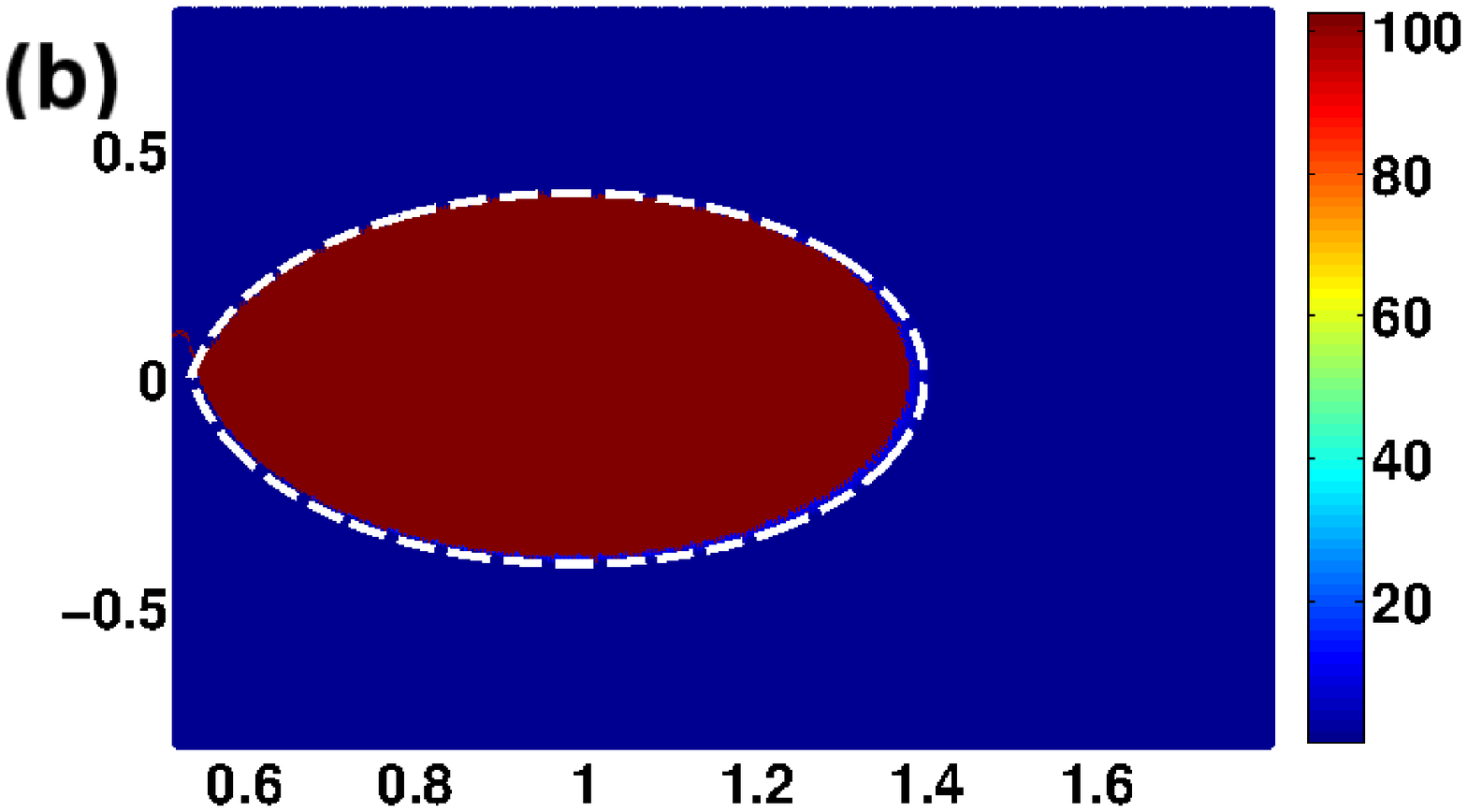}}		
		\centerline{
		\includegraphics[width=0.48\textwidth]{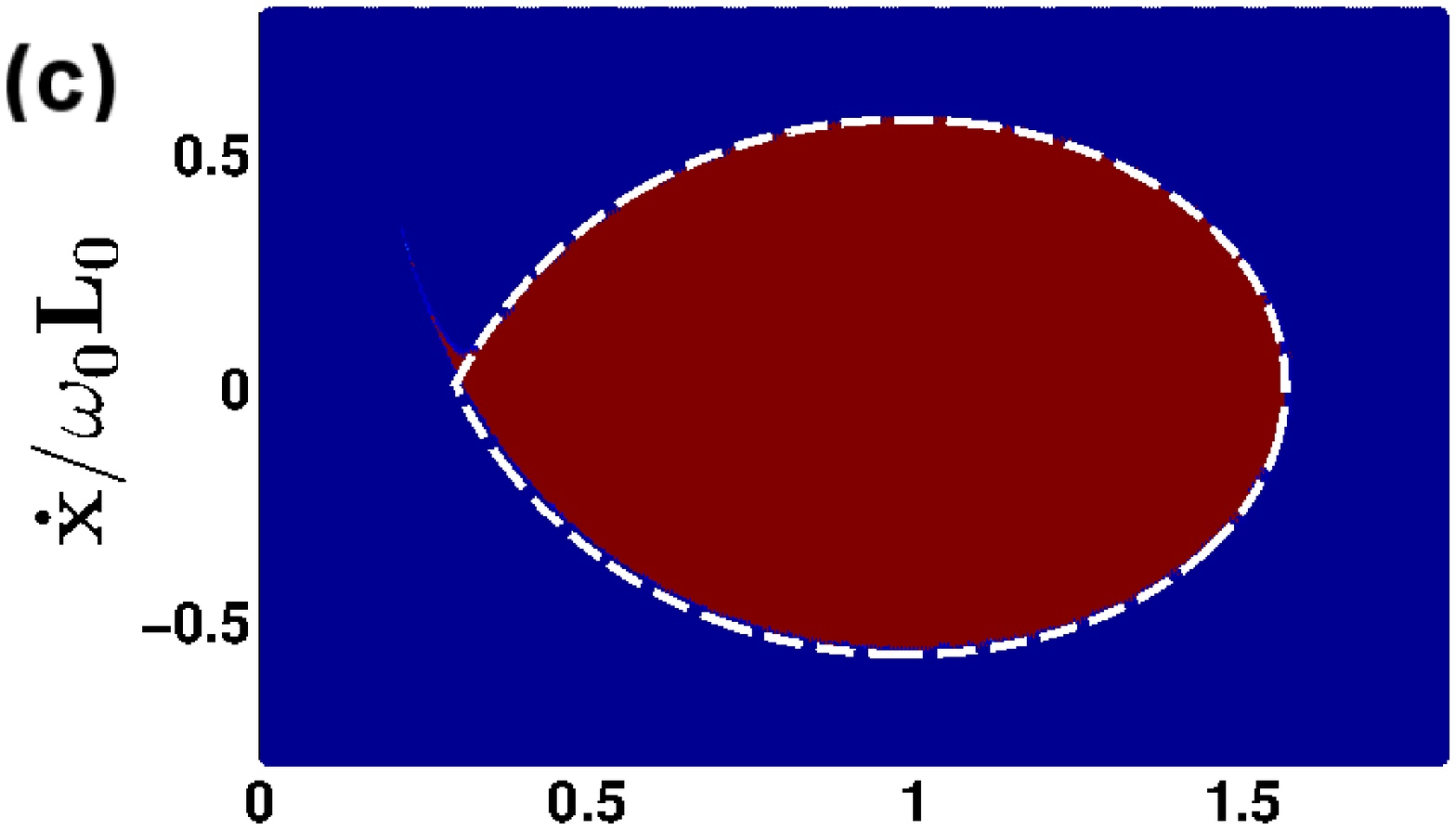}			
		\includegraphics[width=0.48\textwidth]{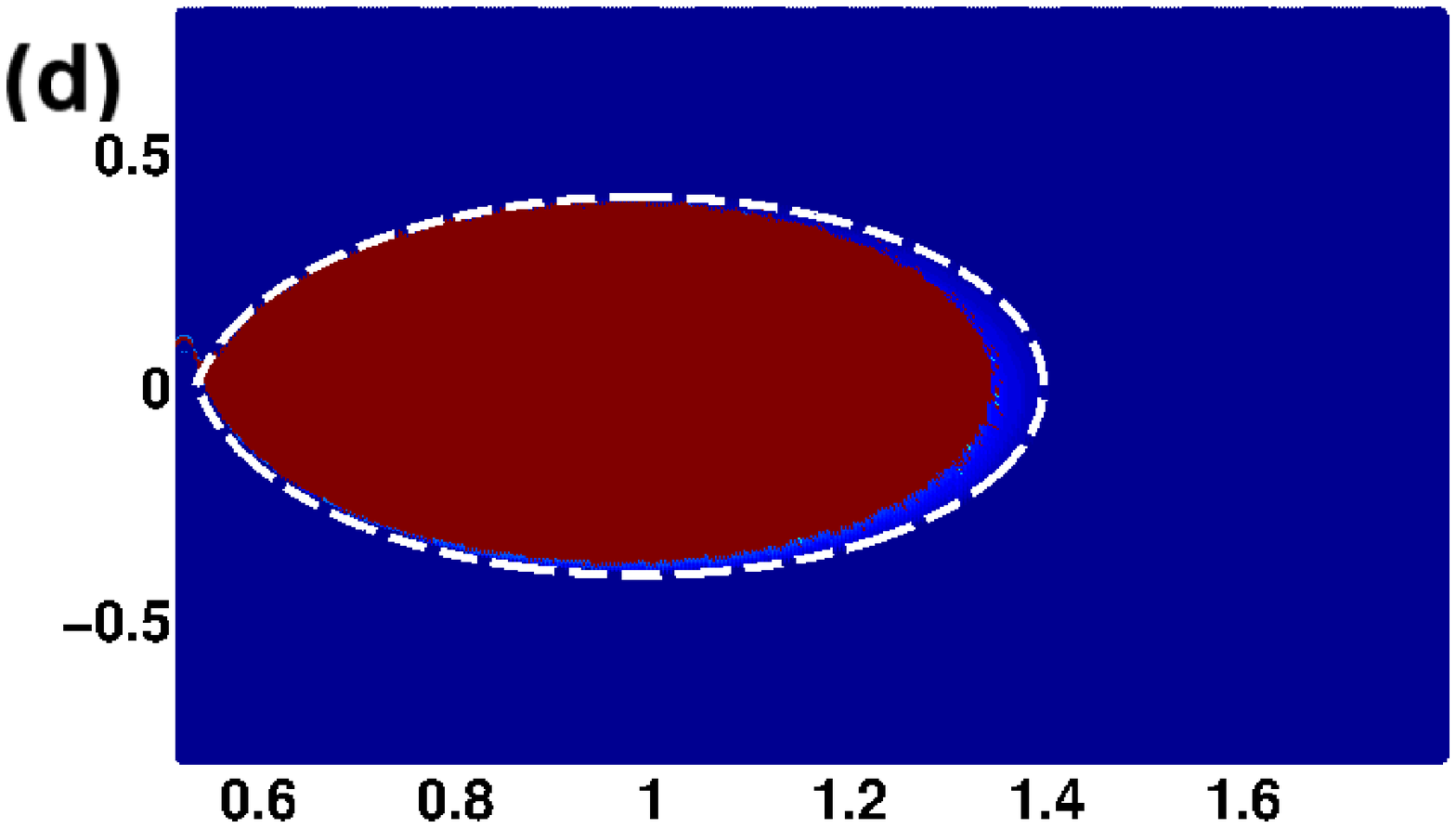}}		
		\centerline{\includegraphics[width=0.48\textwidth]{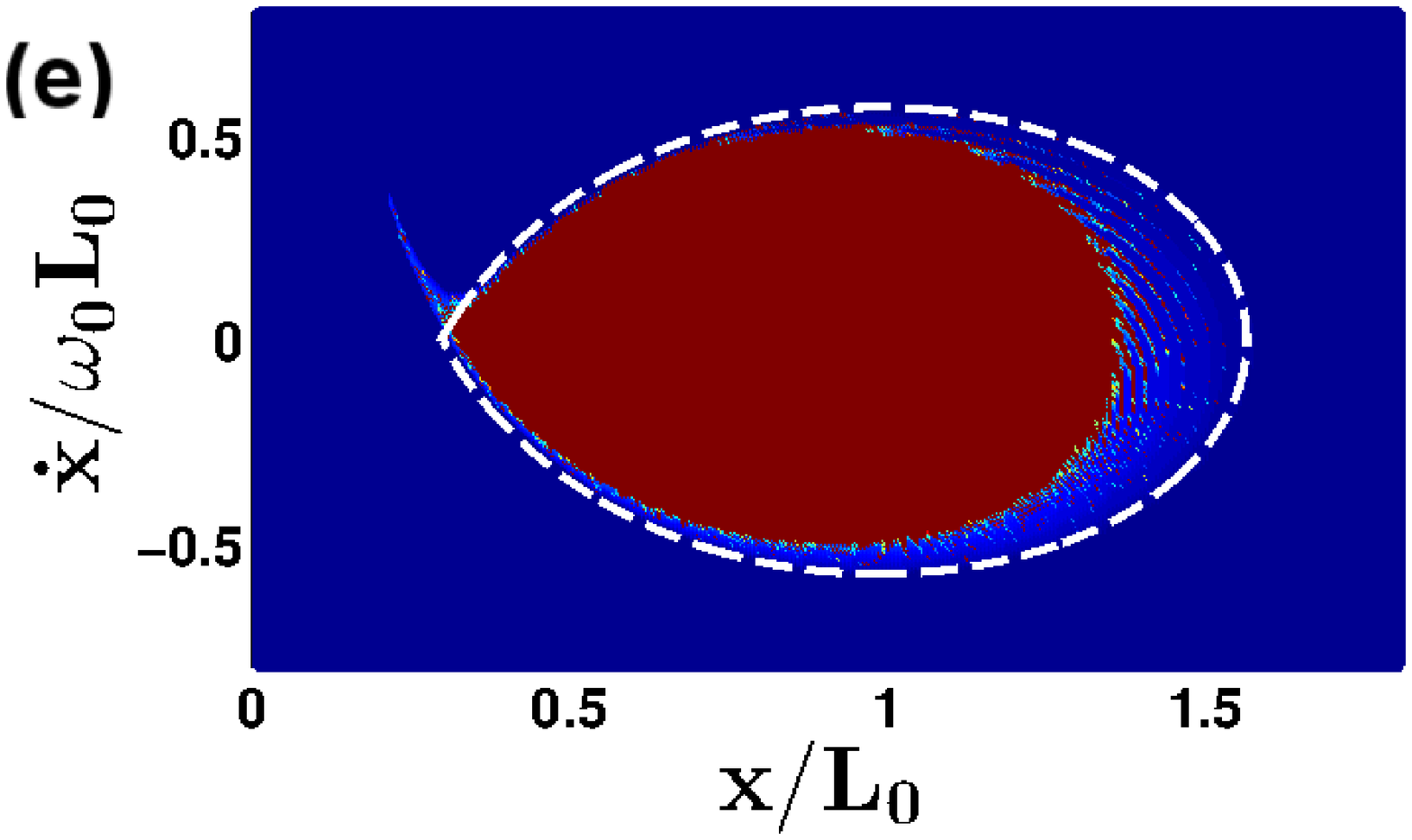}
		\includegraphics[width=0.48\textwidth]{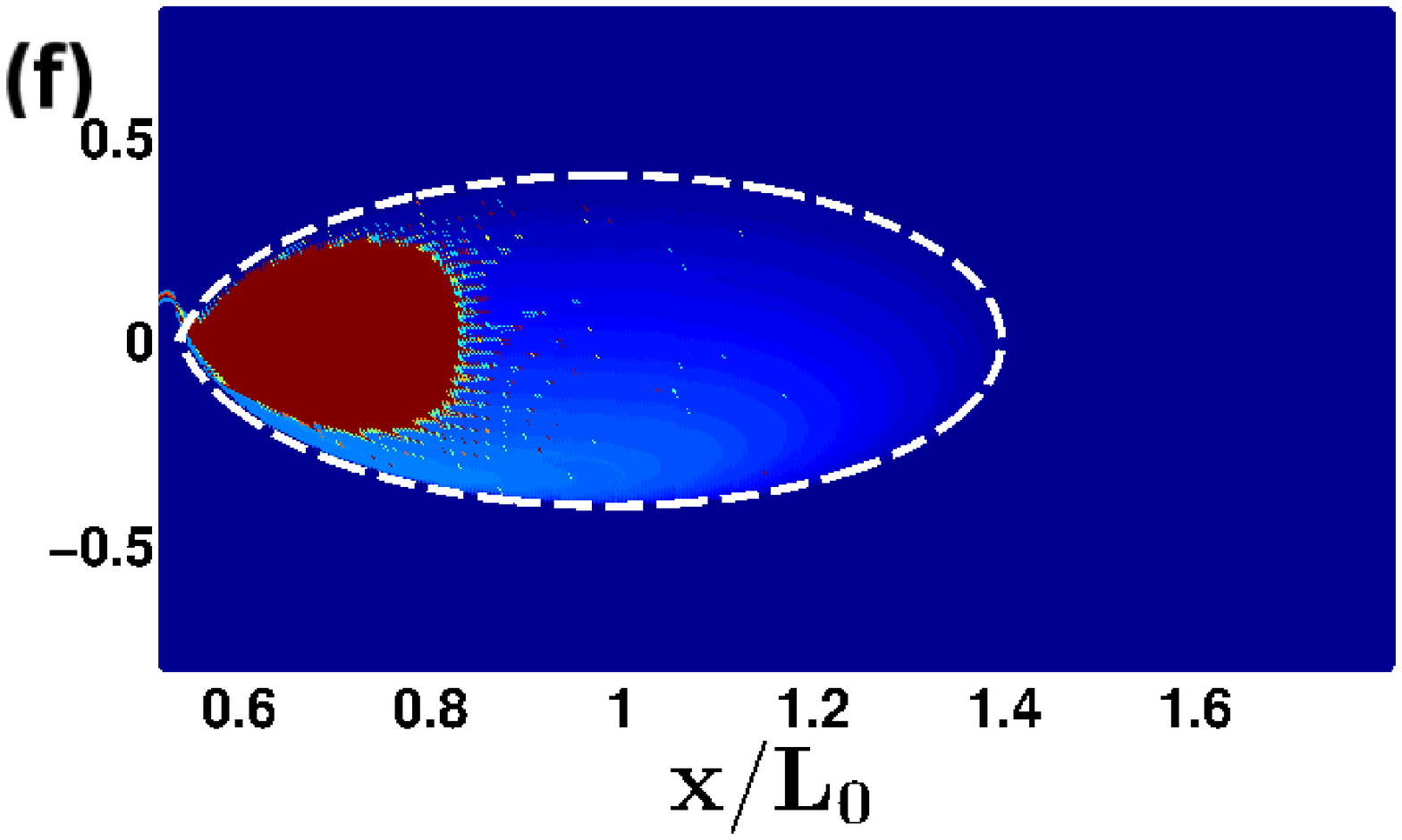}}
										
		\caption{ Grid of 300 $\times$ 300 initial conditions for the system of Eq. \eqref{eq:EOM} for $\epsilon
		=1$. The color bar indicates how much time elapses until stiction occurs, in units of $2\pi/\omega$, with a maximum of 100. The dashed (white) line in each figure shows the homoclinic orbit of the conservative system. The left panels show the results for flat surfaces, whereas the right panels represent rough surfaces. The values of the relevant parameters are: (a) flat, $\alpha=0.03$ (not chaotic); (b) rough, $\alpha=0.03$ (not chaotic); (c) flat, $\alpha=0.02$ (not chaotic); (d) rough, $\alpha=0.02$ (chaotic);   (e) flat, $\alpha=0.005$ (chaotic);  (f) rough, $\alpha=0.005$ (chaotic). In each panel $\omega/\omega_0=1.05$.}
	\label{fig:basin}
\end{figure*}

The results of these computations are shown in Fig.~\ref{fig:basin}. The red areas in Fig.~\ref{fig:basin} indicate the initial conditions for which stiction is avoided for at least 100 periods of the driving. For initial conditions within the dark blue regions, stiction typically occurs within one period. In the conservative case, which is non-chaotic, the border between the blue and red regions coincides with the homoclinic orbit. We study the occurrence of chaotic motion in terms of sensitive dependence of the motion on its initial conditions. In the context of the plots in Fig.~\ref{fig:basin}, this means that there is a region of initial conditions where the distinction between qualitatively different solutions is not clear. Chaotic motion is therefore identified by a lack of a simple, smooth border between the red and the dark blue regions. It should be kept in mind that the finite number of grid points can also make this shape less smooth, which is a numerical artefact not to be confounded with chaotic motion. In the absence of chaotic motion, stiction either occurs on the time scale of one period $2\pi/\omega$, or not at all. However, if the motion is chaotic, stiction may still occur after several tens of periods. Hence chaotic motion concerns also the long term stability of the device.

The two panels at the top, Figs. \ref{fig:basin} (a) and (b), display the results for parameter values above all the threshold curves (Fig.~\ref{fig:threshold}). The Melnikov method predicts that the separatrices will not intersect and hence that the motion of the oscillator will not be chaotic in this case. Indeed, only red and dark blue regions occur, and the border between them is smooth (within numerical accuracy).  It can be seen that the driving pushes the red area to the left of the homoclinic orbit (the dashed white line).This is more pronounced in the rough case (panel (b)) than in the flat case (panel (a)). Also, the damping allows a few initial conditions to the left of the homoclinic orbit to lead to stable actuation.

Panels \ref{fig:basin} (c) and (d) show the results for actuation parameter values in the red region in Fig.~\ref{fig:threshold}. This means that the Melnikov method predicts chaos in the rough case, but not in the flat case. The flat case (panel (c)) is indeed clearly not chaotic: Only dark blue and red areas can be seen. The border between them almost entirely coincides with the homoclinic orbit (the dashed white line). The rough case (panel (d)) is not quite as obvious here. At first sight, it appears not to be chaotic either. However, upon closer inspection it can be seen that the colour of the area between the red region and the dashed line is lighter than the colour of the region outside the homoclinic orbit. This is a (subtle) indication of chaotic motion. In terms of the separatrices, this could mean that they do intersect transversely, but that the angle of intersection is quite small. Something similar can be seen in the case of the Duffing oscillator for parameter values just below the threshold curve (see supplemental material \cite{BroerSupplemental2015} at [URL will be inserted by publisher] for an example of a computation of separatrices of a Duffing oscillator).

Finally, panels \ref{fig:basin} (e) and (f) show the results for actuation parameter values in the blue region of Fig.~\ref{fig:threshold}. Here the motion should be chaotic both for flat and for rough surfaces. This indeed turns out to be the case. In both the flat (panel (e)) and the rough case (panel (f)) stiction can occur after several tens of periods of the driving.

Chaotic motion of Casimir oscillators introduces a considerable risk of stiction.  From a practical viewpoint, chaotic motion should be avoided, since it constitutes uncertainty about the qualitative nature of the solution of the equation of motion. In a chaotic system there is in practice no way to tell whether stiction or stable actuation will occur. Note that in the conservative case, which is non-chaotic, initial conditions to the left of the saddle  equilibrium will \textit{always} lead to stiction. However, this is unrelated to chaos, which is associated with \textit{uncertainty} about the qualitative nature of the solutions.

We have shown that the results of the analytical computations of the Melnikov method (Eq. \eqref{eq:threshold_final} and Fig.~\ref{fig:threshold}) are consistent with the numerical results shown in Fig.~\ref{fig:basin}. We tested our numerical procedure also for the reference system given by the Duffing oscillator for which the threshold curves can be computed analytically and the separatrices can be obtained directly. (See supplemental material at \cite{BroerSupplemental2015} [URL will be inserted by publisher] for a comparison between analytical and numerical threshold curves.)

\section{Conclusions}
We have investigated under what conditions chaotic motion occurs in a damped driven Casimir oscillator. We have demonstrated that the nonlinearity of the Casimir force can give rise to chaotic motion of MEMS at separations below 100 nm. In terms of MEMS applications, chaos can be interpreted as the `blurring' of the distinction between initial conditions leading to stable actuation and the ones leading to stiction. Such uncertainty about the qualitative nature of the motion is highly undesirable for MEM systems. Surface roughness makes the MEM system more susceptible to this effect. Therefore for MEMS applications, it is recommended to minimize surface roughness in order to increase the range of actuation parameter values for which chaotic motion is avoided. Note that, for a flat surface, more initial conditions to the left of the saddle are possible, which always lead to stiction. However, this is unrelated to chaos. It has been established that the homoclinic solution of the conservative system determines the separation range where chaotic motion may occur. This is also the range where the Casimir force, which constitutes the nonlinearity of the equation of motion, is larger for rough surfaces than for flat surfaces.

{
The  method presented in this paper is not restricted to Casimir oscillators.  In fact the Melnikov method is a standard and widely used method for proving the occurrence of chaotic motion. It applies to  basically any periodically  perturbed oscillator that  possesses a homoclinic orbit in the limit of vanishing perturbation. Technically the method  requires the existence of periodic solutions inside of the loop formed by the homoclinic orbit where the period needs to go monotonously  to infinity when the  periodic solutions approach the homoclinic loop. This is however usually the case inside of a  homoclinic loop and hence no restriction. 
For the technical details and further applications, we refer to \cite{GuckenheimerBook} which is a main reference for the Melnikov method. 

}

Furthermore, it is worth emphasizing that {this method} does not require an analytical expression for the Casimir force, or for any other surface force one may want to consider. For example, it could accommodate theoretical roughness corrections to hydrodynamic \cite{Craig2003,*Vinogradova2006} and capillary forces \cite{PeterCapillary} for devices operating in ambient conditions. Other future investigations may be focused on the correlation between the effects of material properties and roughness on the Casimir force.

\begin{acknowledgments}
We gratefully acknowledge useful discussions with H. W. Broer, A. E. Sterk and V. Kraj\~{n}\'{a}k. The authors benefited from exchange of ideas within the network CASIMIR of the European Science Foundation (ESF).
\end{acknowledgments}

\bibliography{Casimir}
\end{document}